\documentclass[12pt]{iopart}

\usepackage{iopams}
\usepackage{graphicx}
\usepackage{color}
\begin{document}

\title{Room temperature triggered single-photon source in the near infrared}

\author{E Wu$^{1,2}$, J R Rabeau$^3$, G  Roger$^4$, F Treussart$^1$, H Zeng$^2$, P~Grangier$^4$, S~Prawer$^5$ and J-F Roch$^1$}

\address{$^1$ Laboratoire de Photonique Quantique et Mol$\rm\acute{e}$culaire, UMR CNRS 8537, Ecole Normale Sup$\rm\acute{e}$rieure de Cachan, France}
\address{$^2$ State Key Laboratory of Precision Spectroscopy, East China Normal University, Shanghai 200062, China}
\address{$^3$ Department of Physics, Macquarie University, Sydney, New South Wales 2109, Australia}
\address{$^4$ Laboratoire Charles Fabry de l'Institut d'Optique, UMR CNRS 8501, Palaiseau, France}
\address{$^5$ Centre of Excellence for Quantum Computer Technology and Quantum Communications Victoria, School of Physics,
University of Melbourne, Victoria   3010, Australia}
\ead{roch@physique.ens-cachan.fr}
\begin{abstract}
We report the realization of a solid-state triggered single-photon source with narrow emission in the near infrared at room temperature. It is based on the photoluminescence of a single nickel-nitrogen NE8 colour centre in a chemical vapour deposited  diamond nanocrystal. Stable single-photon emission has been observed in the photoluminescence under both continuous-wave and pulsed excitations. The realization of this source represents a step forward in the application of diamond-based single-photon sources to Quantum Key Distribution (QKD) under practical operating conditions.
\end{abstract}

\pacs{42.50.Dv, 81.05.Uw, 78.67.Bf}
\submitto{\NJP}

\maketitle

\section{Introduction}
Practical and reliable triggered single-photon sources have attracted a lot of attention in recent years~\cite{NJP_special_issue_SPS}, owing to their numerous applications in quantum information processing. 
Among them, faithful implementation of the BB84 quantum key distribution protocol~\cite{BB84} relies on the use of single-photon pulses which ensure perfect security, over long distances, against eavesdropper attack~\cite{LutkenhausPRA99}. 
As production of single-photon pulses at a sufficiently high rate became practical enough, quantum key distribution experiments was demonstrated with triggered single-photons~\cite{Waks_Nature02,Beveratos_QKD_PRL, Alleaume_NJP} and with heralded single-photons from parametric down-conversion~\cite{Romain_AIP04,Soujaeff_OptExp07}.

Alternate protocols using weak coherent pulses have been proposed~\cite{Scarani_PRL04,Inoue_PRA05,Lo_PRL05}. These protocols, e.g. using ``decoy-state'' are promising alternatives to single-photon QKD in the prospect of increasing the distance of perfectly secure communication compared to the ``classical'' limit that was determined in Ref.\cite{LutkenhausPRA99}. 
It was also recently shown that a modified decoy-state protocol using single-photon pulses can improve practical QKD~\cite{Karlsson_EPL07,Horikiri_PRA06}. This result is another motivation of developing QKD dedicated practical single-photon sources.

In this context, optically-active defects in diamond provide numerous possibilities. Thanks to a near perfect room temperature photostability, the Nitrogen-Vacancy (N-V) colour centre in diamond has been demonstrated as a remarkable source for emitting single-photon pulses on demand~\cite{Beveratos_EPJD}. 

This system successfully led to the realization of the single-photon QKD testbeds under realistic operating conditions~\cite{Beveratos_QKD_PRL, Alleaume_NJP} and to the observation of single-photon interference in the delayed-choice regime~\cite{Jacques_Science}. The NV colour center
has nevertheless a broad spectral emission at room temperature (100~nm FWHM) due to phonon broadening. Spectral filtering is a possible solution but it greatly decreases the effective single-photon transmission rate.

More recently, nickel-nitrogen-related point defects in diamond generated strong interest. These defects can be found in some natural type II-a diamonds and more generally in high-pressure high-temperature (HPHT) grown diamond samples where nickel is used as a solvent/catalyst for the  HPHT crystal growth. Nickel then incorporates in the lattice as a substitutional or an interstitial impurity~\cite{Ni_Yellisseyev}.  In the presence of nitrogen impurities, it can form a variety of nickel-nitrogen colour centres where a single nickel atom is surrounded by different number of nitrogen atoms located at adjacent lattice sites~\cite{Ni_Johnston}.

The photoluminescence of individual nickel-nitrogen-related colour centres has several striking features~\cite{Gaebel_NJP, Wu_OptExpress}: a narrow emission band  around 800~nm almost entirely concentrated in the zero phonon line (ZPL), with a Debye-Waller factor of approximately 0.7 at room temperature and a spectral width of the order of 1~nm; an  excited-level lifetime around 2~ns; and a linearly-polarized light emission.
These centres can be isolated at the single-emitter level in bulk diamond samples~\cite{Wu_OptExpress} but  the high refractive index of diamond leads to a low collection efficiency due to total internal reflection and spherical aberrations. 
However, if the size of the diamond sample is much smaller than the radiated light wavelength, refraction effects can be neglected and the colour centre can be approximated  to a point source radiating at the air-glass interface~\cite{Beveratos_PRA}.
It was demonstrated that nickel-nitrogen-related  centres can be fabricated  by chemical vapour deposition (CVD) of diamond~\cite{Rabeau_APL_NE8}. The control of the CVD process parameters enables one to grow well isolated nanodiamonds with a specific size and with controlled defect  concentration~\cite{Rabeau_NanoLett}, contrary to natural diamond in which the defects are randomly distributed. 
This capability opens numerous possibilities for the development of highly efficient diamond-based single-photon sources, e.g.  by   growing photoluminescing  diamond nanocrystals on an optical fibre endface~\cite{Rabeau_APL_Fiber} or by coupling them  to  microcavity resonance modes~\cite{Butler_APL_Microdisk}.

We report  the observation of individual nickel-nitrogen-related colour centres in well isolated CVD diamond nanocrystals with photoluminescence centered at 793~nm which corresponds to the NE8 colour centre emission~\cite{Zaitsev_TheBook}.
Using a tunable continuous-wave titanium-doped sapphire (Ti:Sa) laser, we measure the photoluminescence excitation (PLE) spectrum of these defects. We then find out the optimal excitation wavelength which leads to the highest signal-to-background ratio with the detection setup of the  experiment.  We then switch to a pulsed excitation regime, using a ps-pulsed laser tuned at this optimal wavelength. Contrary to previous observations on NV colour centers~\cite{Dumeige_JLumin04}, we observe that cw photoluminescence properties of NE8 colour centers, e.g. its photostability, remain unaffected under this pulsed excitation.
Finally, we demonstrate a room-temperature triggered single-photon source based on the time control of the optical excitation adapted to individual NE8 colour centres. This source has all the appropriate features for operating an open-air QKD testbed under daylight.

\section{Material preparation}
The process parameters of diamond synthesis using the CVD technique  enables control over nanodiamond
density and size and also dopant concentration \cite{Rohrer_DRM, Lifshitz_APL}. In a previous work
\cite{Rabeau_APL_NE8}, continuous films of polycrystallite diamond containing single nickel-nitrogen-related  colour
centres were fabricated by CVD technique. Using the same procedure  while adjusting the growth time,
we have prepared well isolated nickel-doped diamond crystallites which have nanometer scale.

These diamond nanocrystals were grown on a  $170$-micron-thick quartz glass
substrate. After ultrasonic nucleation in  a suspension
containing fine nickel powder and diamond powder  with size smaller than $100$ nm,
the substrates  were washed with  acetone and
deionized water and then  loaded into a $1.2$ kW
microwave plasma CVD reactor (ASTeX) for a 30 minutes growth time. The
chamber gas flow ($500$ sccm) consisted of $0.7\%$ CH$_4$ in H$_2$
gas mixture  and the chamber pressure  was maintained at
$40$ mbar during the growth process, with a  substrate temperature
of  $700\,^\circ$C.
Nitrogen was not
deliberately added to the gas feedstock; however,  it was known to be
present at a background level of approximately $  0.1\%$ corresponding  to an N/C atomic ratio $ \approx\! 0.15$.
Due to growth rate estimated to be less than $1$ $\mu$m/h and a deliberately low nucleation density,
isolated nanodiamonds were formed on the substrate surface instead of a polycrystalline thin film.

\begin{figure}[htbp]
\begin{center}
\includegraphics[width=1\textwidth]{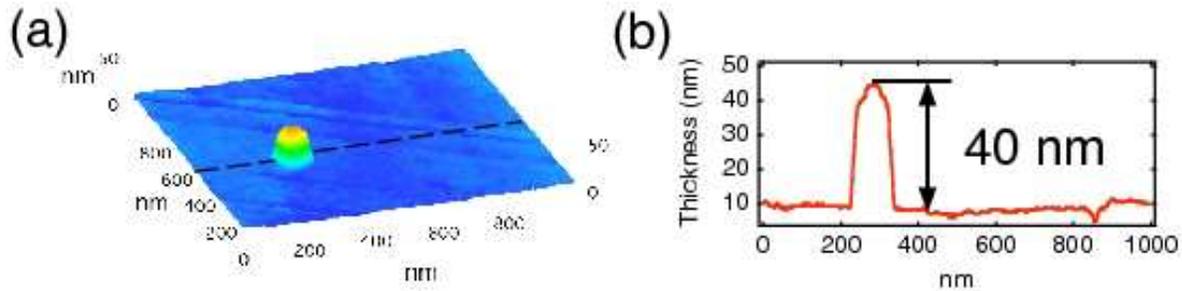}
\caption{{\bf(a) $1$$\times$$1$ $\mu$m$^2$ AFM raster scan
displaying a single diamond nanoparticle grown on the quartz substrate.
 (b) Cross-section of the crystallite indicating a height of approximately 40 nm. The corresponding in-plane dimensions are larger due to the convolution with the radius of curvature of the AFM tip. }}
\label{AFM_Image}
\end{center}
\end{figure}

The sample topography was characterized  using an atomic force microscope
(AFM) in order to evaluate the size and distribution of the
diamond nanocrystals. The surface density was of the order of  $0.1 \, {\rm particle}/(\mu{\rm m})^2$
and the typical nanodiamond size was ranging between 30 nm to 100 nm (figure \ref{AFM_Image}).

\section{Experimental setup}
\begin{figure}[htbp]
\begin{center}
\includegraphics[width=0.65\textwidth]{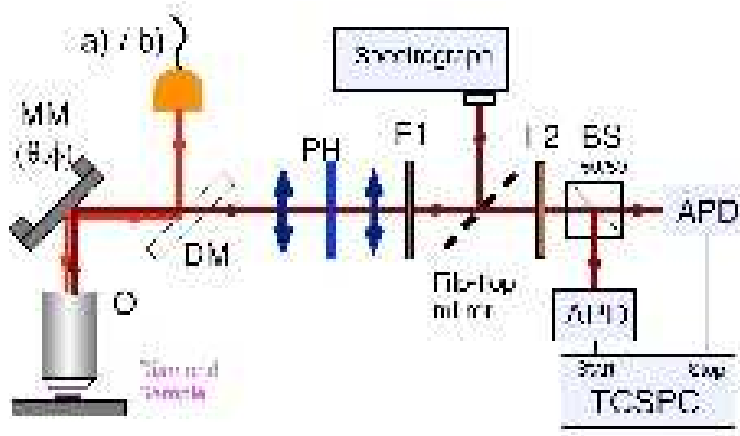}
\caption{{\bf Experimental set-up. Excitation source: a) cw Ti:Sa
laser   tunable from $\bf700$ to $\bf 800$ nm; b) picosecond pulsed
laser emitting at  765~nm with $\bf6$ ps pulse duration and $\bf20$ MHz
repetition rate; MM: Mobile mirror mounted  on a scanning
piezoelectric transducer; O: Microscope objective; DM: Dichroic
mirror; PH: Pinhole with $\bf100$ $\bf{\mu}$m diameter; F1:
Long-pass interference filter transmitting  wavelengths above $\bf{780}$
nm;   F2: Band-pass filter centered at 800 nm and with a bandwith of $\bf10$ nm FWHM;
BS: non-polarized beamsplitter; APD: Silicon avalanche photodiode operated  in photon counting
regime; TCSPC: Time-correlated single-photon counting system
(PicoHarp $\bf300$, PicoQuant, Germany);  Spectrograph:  \it{\bf Micro HR}
\bf(Horiba Jobin Yvon, France) equiped with a   CCD detector cooled at $\bf200$~K and a  $\bf1200$ lines$\bf/$mm grating blazed at $\bf630$ nm. The    spectral   resolution of the spectrograph is approximately $\bf0.25$ nm.}} \label{setup}
\end{center}
\end{figure}

The experimental setup for the optical measurements  of the single nickel-nitrogen-related colour
centres was based on a home-made confocal microscope  (figure \ref{setup}).
 The excitation laser, being either a tunable cw Ti:Sa laser or a picosecond pulsed laser emitting at 765 nm,
   was focused
on the sample with a metallographic microscope objective (O) of
high numerical aperture (${\rm NA} \!=\!0.95$, $\times 100$) which yielded  a spot
size of about $1$ $\mu$m.
By driving a piezo-electric transducer-mounted mirror (MM), the objective allowed for an $x-y$
raster scan of the sample.
Focused   $z - {\rm scan}$  was obtained by
translating the microscope objective with another piezoelectric
transducer.

The same microscope objective also collected the
photoluminescence  from the colour centres  which was then passed
through a dichroic mirror (DM) and optical
filters (F$1$ and F$2$) that  removed any residual excitation
light. Proper     spatial filtering was obtained by focusing in  a pinhole (PH). The transmitted   photoluminescent light  was sent to the
detection arms for spectral characterization and   assessment  of the emitter unicity   by   the
observation of antibunching   using a standard Hanbury Brown and Twiss   (HBT)  setup for photon
correlation measurements \cite{Brouri_OpticsLetters,Kurtsiefer_PRL}.

\section{Continous-wave excitation}
\begin{figure}[htbp]
\begin{center}
\includegraphics[width=0.8\textwidth]{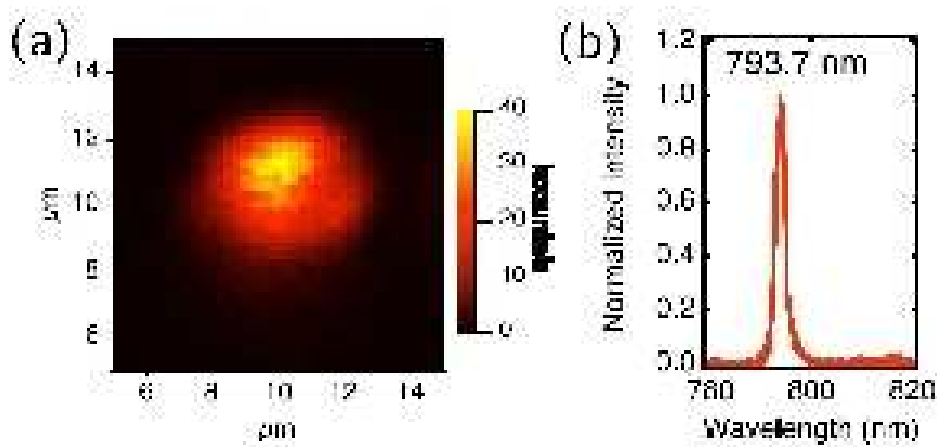}
\caption{{\bf (a) Confocal microscopy raster scan over a photoluminescent  nanodiamond under cw excitation at 765 nm. The colour scale
indicates the total photoluminescence intensity at the output of the   two detectors
  of  the HBT setup. The peak
counting rate  value is approximately $70$~kcounts/s while the background level is of the order of
$0.8$~kcounts/s. (b) Corresponding
photoluminescence spectrum of the single emitter displaying the characteristic emission line of the NE8 colour centre.
}} \label{scan}
\end{center}
\end{figure}

A raster scan over a photoluminescent  nanodiamond is shown in figure \ref{scan}(a).
The bright spot gave rise to a sharp spectral emission line centered at 793.7 nm with a width of  2-nm FWHM,
 as displayed in figure \ref{scan}(b).
This feature is characteristic of the NE8 nickel-nitrogen colour centre \cite{Zaitsev_TheBook}.
 The atomic structure of this point defect was suggested to consist of a nickel atom located at equal distance of two carbon vacancies and associated to four nitrogen impurity atoms in its first coordination shell
 \cite{Ni_Nadolinny}.

\begin{figure}[htbp]
\begin{center}
\includegraphics[width=0.6\textwidth]{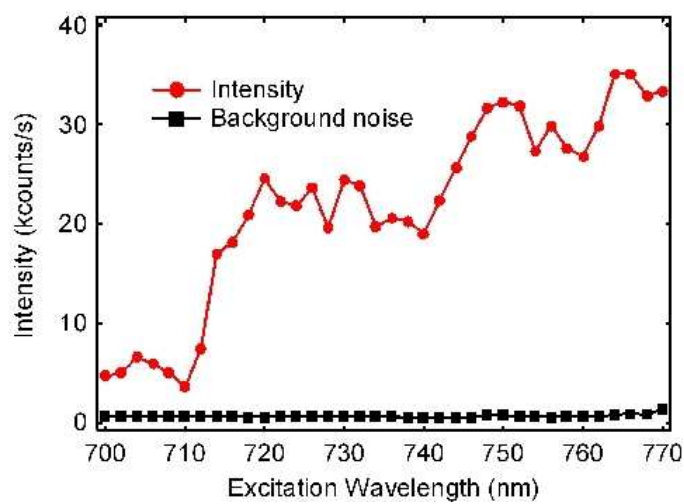}
\caption{{\bf 
Photoluminescent excitation   spectrum of the single NE8 colour centre associated to figure  \ref{scan}.
  The counting rate  corresponds to the sum of the output of the
  two detectors of the HBT setup.   Red dots and black squares respectively represent
the evolution of the total counting rate and of the background as a function of the excitation wavelength,
while maintaining  a  constant excitation power of 1~mW focused on the sample.}}
\label{excitation}
\end{center}
\end{figure}

Photoluminescence of this NE8 colour centre was then recorded
as a function of the cw excitation wavelength, tuned from 700~nm
to 770~nm. Both the position and the shape of the emission line  at $793.7$ nm  remained identical.
As shown in figure~\ref{excitation},  the  photoluminescence intensity increased with the excitation
wavelength    while the background level
remained constant except for a small increase when the excitation
wavelength reached  the cutoff edge of the long-pass filter F1.
The highest signal-to-background ratio of  $ \approx\!  80$ was achieved for a
765 nm excitation wavelength, which corresponds to a second-harmonic frequency
of the $1.55\, \mu{\rm m}$ telecom band.

 \begin{figure}[htbp]
\begin{center}
\includegraphics[width=0.65\textwidth]{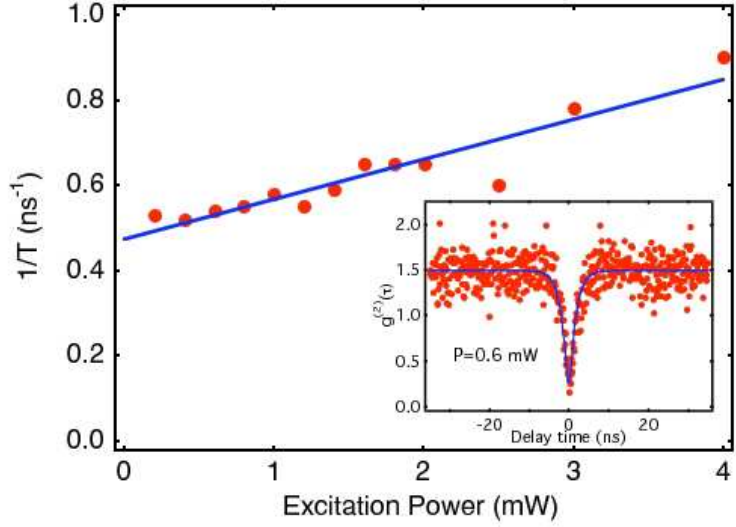}
\caption{{\bf Evolution of the emission rate  $\bf{1/{\it T}}$ (red dots) as a
function of the excitation power and associated  linear fit (blue line).
Inset: Normalized normally-ordered intensity autocorrelation function
$\bf{g^{(2)}(\tau)}$ associated to a time bin of $\bf 0.128$ ns. The histogram of time intervals between successive  detection events (red dots) has been  recorded for a  $\bf{0.6}$ mW  excitation power and a $\bf{800}$ s total integration time. The corresponding detectors counting rates are $\bf{15.8}$ and  $\bf{18.2}$ kcounts/s. The solid
blue line is the intensity autocorrelation function computed for a single emitter with random on/off emission, convoluted with the  instrumental response function of the photon correlation measurement setup (see Ref.\cite{Wu_OptExpress}). }}
\label{fig_lifetime}
\end{center}
\end{figure}

The HBT setup records  the histogram of the time
separation  between successive photodetection events. 
For our operating conditions and at the limit of  short time scale, this histogram is
equivalent to a measurement of the normally-ordered  intensity autocorrelation function $g^{(2)} (\tau) =\langle \, : \!\! I(t)I(t+\tau) \!\! : \,  \rangle/\langle \, : \!\! I(t)  \!\! : \, \rangle^2$. As   shown
 in the inset of figure \ref{fig_lifetime}, a  distinct minimum at zero delay,
  corresponding to  $g^{(2)}(0) \! \approx \! 0.2$,   is observed.
This indicates that the addressed colour centre   was indeed   a single emitter.
As explained in Ref.~\cite{Wu_OptExpress}, fits of    $g^{(2)} (\tau) $ recorded with increasing  excitation powers  yield
the evolution of the exponential decay time $T$  of the emitter  as it gets more and more saturated (figure \ref{fig_lifetime}).
 The extrapolated value for a null excitation intensity allows one to infer an emission lifetime of
approximately 2~ns. This value is similar to the one that has been previously measured  for individual nickel-nitrogen-related colour
 centres   in  a type II-a
natural diamond sample  \cite{Wu_OptExpress}.

\section{Pulsed excitation}
The pulsed excitation starts from a mode-locked Erbium-doped fibre laser (PriTel, USA). The laser output,  at a wavelength of  1530 nm  and emitting    a train of a 6.1~ps pulses with a 20~MHz repetition rate, is then amplified in a fiber amplifier and frequency doubled in a periodically-poled lithium niobate  crystal. We then obtained picosecond pulses of energy 0.75 nJ at the optimized wavelength of 765 nm.

 The TCSPC unit previously used to record the intensity autocorrelation function in the cw excitation regime  was employed to  measure the
  histogram of  time delays
between the laser pulses that trigger the  colour centre photoluminescence and  the corresponding photon detection events.
An exponential decay fit yields an emission lifetime of 2.1~ns, in agreement with the value  inferred from
the antibunching curves recorded under   cw excitation for the same wavelength
 (see  figure~\ref{fig_lifetime}).

\begin{figure}[htbp]
\begin{center}
\includegraphics[width=0.65\textwidth]{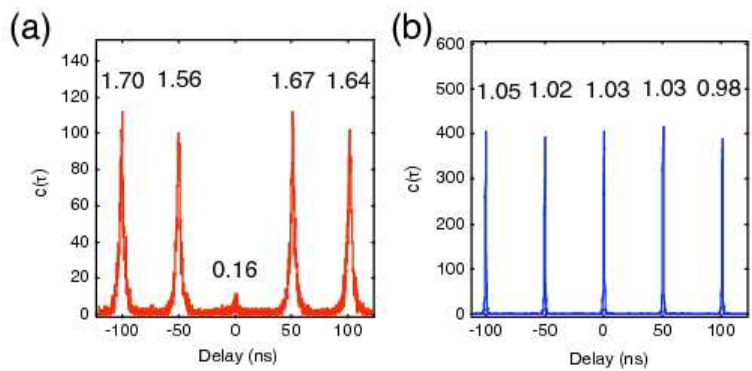}
\caption{{\bf Normally-ordered intensity autocorrelation functions of (a) a single NE8 colour centre under a pulsed excitation, corresponding to an average power of 50~$\mu$W focused on the sample,  and of (b) attenuated pulses
from the excitation laser.  The
normalized total number of coincidence counts associated to each peak  is indicated on its top. }}
\label{g2_pulse}
\end{center}
\end{figure}

The intensity autocorrelation function $g^{(2)} (\tau)$ is then measured under this pulsed excitation regime
and compared to equivalent classical light pulses with Poissonian photon number statistics.  Figure~\ref{g2_pulse}
shows the corresponding data for the triggered emission of the single NE8 defect and  for the attenuated laser pulses, with peaks separated by the excitation
repetition period of $\theta = 50 \, {\rm ns}$  \cite{Brunel_PRL}.
In the case of the single NE8 defect, the peak at $\tau = 0$ has almost disappeared  which implies that the probability
of having two photons in the same output pulse     is greatly reduced compared to the equivalent Poissonian statistics.

To estimate the quality of the single-photon source, we then quantified this reduction following a procedure described in Ref.~\cite{Beveratos_EPJD}.
The total number of coincidence counts $c(m)$ associated to peak $m$ is normalized as:
\begin{equation}
C_N (m) \; = \; \frac{c(m)}{ N_1 \,  N_2 \, \theta \, T_{\rm acq}}
\end{equation}
where $N_1$ and $N_2$ are the counting rates at the output of the two detectors of the HBT setup and $T_{\rm acq}$ is the acquisition time during which     the temporal correlation histogram has been recorded.

The normalized $C_N$ value for each peak is indicated in
figures~\ref{g2_pulse}(a) and \ref{g2_pulse}(b). For   attenuated  pulses from  the excitation laser, $C_N$   is equal to unity  within the measurement uncertainty, as it is expected for Poissonian photon number fluctuations.  For the single NE8 colour centre emission, the peak at $\tau =0$ corresponds to $C_N (0) = 0.16$, well below unity.
The difference between the null value that would correspond to an ideal single-photon source is attributed to a lower signal-to-background ratio observed under pulsed excitation compared to the one measured under cw excitation (see figure~\ref{scan}).

Figure~\ref{g2_pulse}(a) also shows that the peaks corresponding to $m \not= 0$ are associated to a   $C_N$ value larger than unity.
This is attributed to the existence of a ``dark'' metastable state which is also responsible for the bunching effect observed  in the photon correlation measurement under cw excitation.
This bunching corresponds to a limit value of $g^{(2)} (\tau)$ higher than unity when $|\tau|$  exceeds the characteristic decay time $T$  (see the inset of figure~\ref{fig_lifetime}) \cite{Wu_OptExpress}.
Using a model that accounts for  random {\sc on}/{\sc off} emission  \cite{Santori_PRL}, the normalized total number of coincidence counts  for   peak
$m \not= 0$ is given by:
\begin{equation}
C_N (m \neq 0)=1+\frac{T_{\rm off}}{T_{\rm on}} \, {\rm exp} \left\lbrack   -\left( \frac{1}{T_{\rm on}}  + \frac{1}{T_{\rm off}}\right) |m| \theta  \right\rbrack
\end{equation}
where $T_{\rm on}$ is the mean duration during which the emitter is active
and $T_{\rm off}$ the mean duration  during which it is trapped in the dark state and its emission turned off.
Fits of  the $C_N$ values  give  $T_{\rm on} = 9.1\, \mu{\rm  s}$  and
$T_{\rm off} =   7.0\, \mu{\rm  s}$   for  an average excitation power of $50 \, \mu{\rm W}$ focused on the sample.

\begin{figure}[htbp]
\begin{center}
\includegraphics[width=0.5\textwidth]{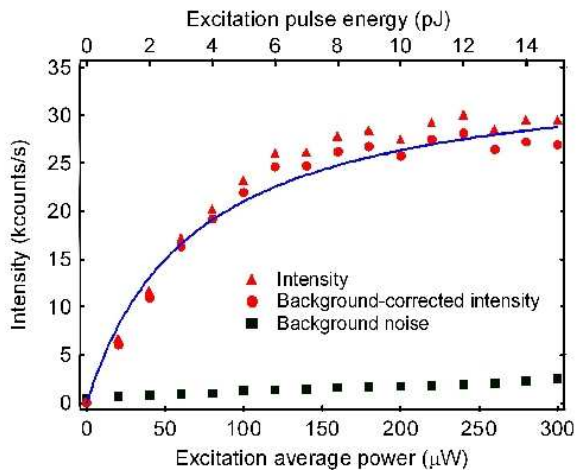}
\caption{{\bf Detection counts rate as a function of the average
excitation power focused on the sample, in the pulsed excitation regime. Experimental data are presented with points (triangle for the   detection counting rate, dots for the background-corrected detected counting rate, and squares for the background level).  The solid line is a fit of the background-corrected signal using equation (\ref{eq_saturation}).}}
\label{saturation}
\end{center}
\end{figure}

These values can be used to estimate the overall efficiency of the single-photon source.
The rate $R$ of total detection counts as a function of the average excitation power
$P$  is shown in figure~\ref{saturation} and can be fitted by
\begin{equation}
R=R_\infty\frac{P}{P+P_{\rm sat}}
\label{eq_saturation}
\end{equation}
where $P_{\rm sat}$ is the saturation average excitation power and $R_\infty$ is
the rate of detection counts at the limit of  saturated emission. Fit of the excitation saturation curve shown in figure~\ref{saturation} yields  $R_\infty =  35$ kcounts/s and $P_{\rm sat}= 66 $~$\mu$W.
Taking into account the quantum efficiency of the emitter resulting from parameters $T_{\rm on}$ and   $T_{\rm off}$  and  the overall detection efficiency $\eta_{\rm det}$  of the setup,  the saturated counting  rate $R_\infty$  can also be  estimated by:
\begin{equation}
R_\infty=\eta _{\rm det} \, \frac{T_{\rm on}}{T_{\rm on}+T_{\rm  off}} \, \frac{1}{\theta} \, .
\end{equation}
The inferred  value of this parameter    allows us
to estimate the   detection efficiency $\eta_{\rm det} \approx\! 0.35\%$ for the single NE8 colour centre in a CVD diamond nanocrystal.
This value is three times larger than the one previously
estimated for the emission of nickel-nitrogen-related colour centres in a bulk diamond sample \cite{Wu_OptExpress}. The
improvement in  the collection efficiency is attributed  to the sub-wavelength size of the nanocrystal structure.

\section*{Conclusion}
NE8 nickel-nitrogen colour centers were observed in CVD-grown diamond nanocrystals. Optimization of the CVD growth technique lead to well separated nanocrystals, in which we could observe single colour centers with a high signal-to-background ratio. The photoluminescence dynamics of NE8 colour centers, previously studied in bulk samples, was demonstrated to scale down to these nanosize structures. Moreover, a higher single-photon detection efficiency compared to the previously reported value in bulk diamond~\cite{Wu_OptExpress} was achieved by avoiding the influence of the high index of refraction of diamond.

Both cw and pulsed excitations were used for single-photon emission by these NE8 colour centers. A triggered single-photon source was realized emitting in the $793\pm 1$~nm transmission window. We determined  the optimal pulsed excitation wavelength according to the excitation spectrum recorded in the cw regime for a single emitter.

We now plan to increase the triggered single-photon emission rate by growing nanodiamonds on optimized photonic structures, and to implement this single-photon source in an open-air QKD testbed.

\ack
We thank Patrick Georges for the loan of the cw Ti:Sa laser and we acknowledge the help of Vincent Jacques. This work was supported by the ``EQUIND''  and the ``Nano4Drugs'' projects  funded by  European Commission (FP6   project numbers IST-034368 and   LSHB-CT-2005-019102),  the ``PROSPIQ'' project funded  by   Agence Nationale de la Recherche,  C'Nano Ile-de-France and
Australian Research Conseil Discovery Projects Scheme. E~Wu and H~Zeng acknowledge financial support from EADS and  Programme Hubert Curien of Minist\`ere des Affaires \'Etrang\`eres.

\end{document}